\documentclass[aps,twocolumn,superscriptaddress,fleqn,showpacs,prl]{revtex4}
\usepackage{amsmath,amssymb}
\usepackage{graphicx}

\begin{document}

\title{Entrainment of randomly coupled oscillator networks by a pacemaker}
\author{Hiroshi Kori} \email[E-mail address:
]{kori@fhi-berlin.mpg.de} \affiliation{Department of Physics,
Graduate School of Sciences, Kyoto University, Kyoto 606-8502, Japan}
\affiliation{Abteilung Physikalische Chemie, Fritz-Haber-Institut der
Max-Planck-Gesellschaft, Faradayweg 4-6, 14195 Berlin, Germany}
\author{Alexander S. Mikhailov} \affiliation{Abteilung Physikalische
Chemie, Fritz-Haber-Institut der Max-Planck-Gesellschaft, Faradayweg
4-6, 14195 Berlin, Germany} \date{20.FEB.2004}

\begin{abstract}
 Entrainment by a pacemaker, representing an element with a higher
 frequency, is numerically investigated for several classes of random
 networks which consist of identical phase oscillators. We find that the
 entrainment frequency window of a network decreases exponentially with
 its depth, defined as the mean forward distance of the elements from
 the pacemaker. Effectively, only shallow networks can thus exhibit
 frequency-locking to the pacemaker. The exponential dependence is also derived
 analytically as an approximation for large random asymmetric networks.
\end{abstract}
\pacs{05.45.Xt, 89.75.Fb, 87.18.Sn}

\maketitle

Pacemakers are wave sources in distributed oscillatory systems typically
associated with a local group of elements having a higher oscillation
frequency. Target patterns, generated by pacemakers, were the first complex
wave patterns observed in the Belousov-Zhabotinsky system \cite{zhab}.
Pacemakers play an important role in functioning of the heart \cite{winfree}
and in the collective behavior of {\em Dictyostelium discoideum} \cite{Lee}.
They are also observed in large-scale ecosystems \cite{Blasius}. In addition
to pacemakers produced by local heterogeneities in the medium \cite{Fife},
self-organized pacemakers in uniform birhythmic media have been
theoretically studied \cite{Stich}. While the majority of related
investigations have so far been performed for systems with local diffusive
coupling between the elements, pacemakers can also operate in oscillator
networks with complex connection topologies. The circadian rhythm in mammals
is a daily variation of 24 hours that regulates basic physiological
processes in such animals \cite{moore97}. It is produced by a complex
network of neurons forming the so-called suprachiasmatic nucleus (SCN) \cite
{abrahamson01}. As recently shown, this oscillator network undergoes
spontaneous synchronization in absence of any environmental input, but its
intrinsic synchronization period is then significantly longer than 24 hours 
\cite{yamaguchi03}. Therefore, the actual shorter rhythm results from the
environmental entrainment and must be externally imposed. The entrainment is
mediated by direct photic inputs from eyes into the SCN, which undergo
periodic daily variation. However, it is known that only a distinct subset
of neurons in this network is directly influenced by photic inputs \cite
{kuhlman03}. Hence, functioning of this particular neural system is
crucially dependent on the ability of the entire complex network to become
entrained by an external pacemaker. Analogous behavior can also be expected,
for example, in heterogeneous arrays of globally coupled electrochemical
oscillators where synchronization and entrainment have been experimentally
demonstrated \cite{Hudson}.

To understand operation of pacemakers in networks with complex connection
topologies, action of a pacemaker in a random oscillator network should
first be investigated. In this Letter, networks of identical phase
oscillators with random connections are considered. A pacemaker is
introduced as a special element whose oscillations have a higher frequency
and are not influenced by the rest of the system. Depending on the pacemaker
frequency and the strength of coupling, the pacemaker can entrain the entire
network, so that the frequencies of all its elements become equal to that of
the pacemaker. We find that the entrainment window decreases exponentially
with the depth of a network, defined as the mean forward distance of its
elements from a pacemaker, and thus only shallow networks can effectively be
entrained. This result is confirmed in numerical simulations for several
different classes of random networks, including small-world graphs. It is
further analytically derived as an approximation for random networks with
asymmetric connections.

We consider a system of $N+1$ phase oscillators, one of them being a
pacemaker. The model is given by a set of evolution equations \cite
{kuramoto84} for the oscillator phases $\phi _{i}$ and the pacemaker phase $%
\phi _{0}$, 
\begin{eqnarray}
\dot{\phi}_{i} &=&\omega -\frac{\kappa }{pN}\sum_{j=1}^{N}A_{ij}\sin (\phi
_{i}-\phi _{j})-\mu B_{i}\sin (\phi _{i}-\phi _{0}),  \nonumber \\
\dot{\phi}_{0} &=&\omega +\Delta \omega .  \label{model}
\end{eqnarray}
The topology of network connections is determined by the adjacency
matrix ${\mathbf A}$ whose elements $A_{ij}$ are either $1$ or $0$.  The
element with $i=0$ is special and represents a pacemaker. Its frequency
is increased by $\Delta \omega$ with respect to the frequency $\omega$
of all other oscillators \footnote{Note that the system (\ref{model}) is
invariant under transformation $\omega \to - \omega, \Delta \omega \to
-\Delta \omega, \phi \to -\phi$, and therefore the same entrainment
behavior takes place when $\Delta \omega <0$.}.  The pacemaker is acting
on a randomly chosen subset of $N_{1}$ elements, specified by $B_{i}$
taking values $1$ or $0.$ The total number of connections to the
pacemaker, $N_{1}=\sum_{i}B_{i}$, is fixed. The coupling between
elements inside the network is characterized by strength $\kappa$.  The
strength of coupling from the pacemaker to the network elements is
determined by the parameter $\mu $. In absence of a pacemaker, such
networks undergo autonomous phase synchronization at the natural
frequency $\omega $.  Without loss of generality, we put $\omega =0$.
Moreover, we rescale time as $t^{\prime }=t$ $\Delta \omega $ and
introduce rescaled coupling strengths $\kappa ^{\prime }=\kappa /\Delta
\omega ,\mu ^{\prime }=\mu /\Delta \omega$. After such rescaling, the
model takes the form of Eqs. (1) with $\Delta \omega =1$ and $\omega =0$
(we drop primes in the notations for the rescaled couplings). In terms
of the original model (1), increasing the rescaled coupling between the
elements is equivalent either to an increase of coupling $\kappa$ or to
a decrease of the relative pacemaker frequency $\Delta \omega $.

The presence of a pacemaker imposes hierarchical organization. For any
node $i$, its distance $h$ with respect to the pacemaker is given by the
length of the minimum forward path separating this node from the
pacemaker. All $N_{1}$ elements in the group directly connected to the
pacemaker have distances $h=1$, the next elements which are connected to
the elements from this group have distances $h=2$, etc.  Thus, the whole
network is divided into a set of shells \cite{dorogovtsev03}, each
characterized by a certain forward distance $h$ from the pacemaker. The
set of numbers $N_{h}$ is an important property of a network. The depth
$L$ of a given network, which is the mean distance from the pacemaker to
the entire network, is introduced as $L=(1/N)\sum_{h}hN_{h}$. It should
be noticed that such ordering of network nodes is based solely on the
forward connections down the hierarchy and does not depend on the
distribution of reverse (upward) connections in the system.

\begin{figure}
\resizebox{8.6cm}{!}{\includegraphics{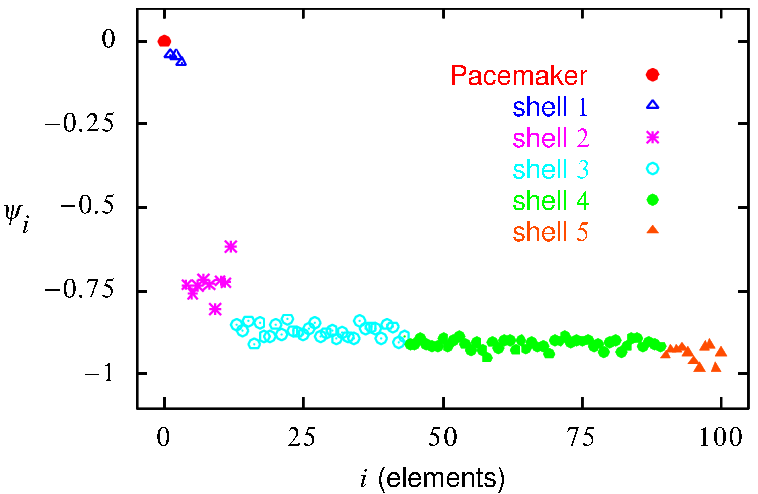}}
\caption{(color online) Phases of elements in the entrained state;
 $N=100$, $p=0.05$, $N_1=3$, $\kappa =100$ and $\mu =1000$.}
\label{fig:phase}
\end{figure}
First, we investigated {\em standard random asymmetric networks}, where
independently for all connections $A_{ij}=1$ with probability $p$ and $
A_{ij}=0$ otherwise. Only sparse random networks with relatively low
mean connectivity $p$ and a small number $N_{1}$ of elements directly
connected to the pacemaker were considered. Numerical simulations were
performed for the networks of size $N=100$ starting with random initial
conditions for the phases of all oscillators. For each oscillator, its
effective long-time frequency $\omega _{i}$ was computed as $\omega
_{i}=T^{-1}\left[ \phi _{i}(t_{0}+T)-\phi_i (t_{0})\right]$ with
sufficiently large $T$ and $t_{0}$. The simulations show that the
response of a network to the introduction of a pacemaker depends on the
strength $\kappa $ of coupling between the oscillators.  When this
coupling is sufficiently large (and coupling $\mu $ to the pacemaker is
also sufficiently strong as assumed below), the pacemaker entrains the
whole network (i.e., $\omega _{i}=1$ for all elements $i$). The frozen
relative phases $\psi_{i}\equiv \phi _{i}-\phi _{0}$ are displayed in
Fig.~1. Here, the elements are sorted according to their hierarchical
shells.  Despite random variations, there is a clear correlation between
phases of oscillators and their positions in the hierarchy.  Generally,
the phase decreases for deeper shells, and the phase difference between
the neighboring shells rapidly becomes smaller as deeper shells are
considered. As the coupling strength $\kappa $ is decreased, the
entrainment breaks down at a certain threshold value $\kappa _{{\rm
cr}}$. Our simulations show that synchronization between the first and
the second shells was almost always the first to break down, and the
frequencies of the second and deeper shells remained equal in most cases
for the considered random networks.

Figure 2 displays in the logarithmic scale the thresholds $\kappa_{\rm
cr}$ for a large set of networks with different depths and different
numbers of elements in the first shell. Each group with a certain
$N_{1}$ is displayed by using its own symbol. Every such group generates
a cluster of data points. Correlation between the entrainment threshold
and the network depth is apparent. The distributions inside each cluster
and the accumulation of the clusters yield the dependence $\kappa_{{\rm
cr}}(L)$ of the entrainment threshold on the network depth. Note that
the statistical variation of the data becomes larger for deeper networks
with larger $L$ and for smaller $pN$. Similar dependence was found for
the networks with different mean connectivity $p$ (see inset). Remarkably,
the observed dependences could be well numerically approximated
by the exponential dependence
\begin{equation}
 \kappa_{{\rm cr}} \propto (1+pN)^{L}.  \label{exponential}
\end{equation}
\begin{figure}[t]
\resizebox{8.6cm}{!}{\includegraphics{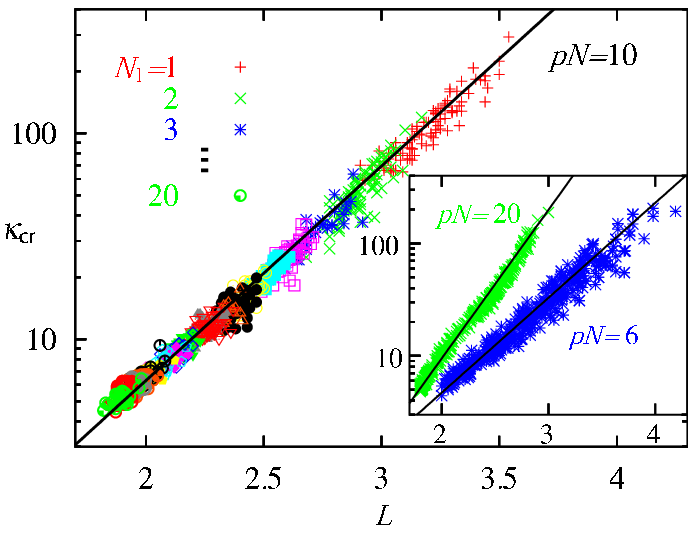}} \caption{(color online)
Dependence of the entrainment threshold on the depth $L$ for an ensemble
of random networks with $N=100$ and $p=0.1$. In the inset, respective
data for networks with $p=0.06$ and $0.2$ is plotted.  Solid lines are
the exponential functions $c_p(1+pN)^L$ with appropriate fitting
parameters $c_p$. } \label{fig:Kc}
\end{figure}

As the second class, {\em asymmetric small-world networks}
\cite{watts98} were considered.  To generate them, we first
constructed a one-dimensional lattice of $N$ elements where each
element had incoming connections from up to its $k$th neighbor (the
degree was thus $2k$). Then a randomly chosen link in the lattice was
eliminated and a distant connection between two independently randomly
chosen elements was introduced. This construction was repeated $qN$
times, with the parameter $q$ specifying the randomness of a network.
When $q$ was small, the network was close to a lattice and, in this
case, we have seen that stable wave solutions with different winding
numbers were possible, depending on initial conditions (cf.
\cite{kuramoto84,manrubia}). To avoid this, we chose almost
synchronized states as initial conditions. The entrainment thresholds
for such small-world networks are displayed in Fig.~3 and again show a
clear correlation between $\kappa _{{\rm cr}}$ and $L$. The dependence
on the depth is approximately linear in lattices ($q=0$), but it
becomes strongly nonlinear even when small randomness is introduced.
For $q=0.1$, the dependence is already approximately exponential,
though the dispersion of data is strong. As randomness $q$ is
increased, the dependence approaches that of the standard random
networks with $pN=2k$.

We have also investigated {\em asymmetric scale-free random networks}
\cite{newman01}, {\em asymmetric regular random networks} (where every
element has exactly the same number of either incoming or outgoing
connections), and {\em symmetric standard random networks}. For all of
them, approximately exponential dependences of the entrainment
threshold on the network depth were observed in a large parameter
region.

\begin{figure}[t]
\resizebox{8.6cm}{!}{\includegraphics{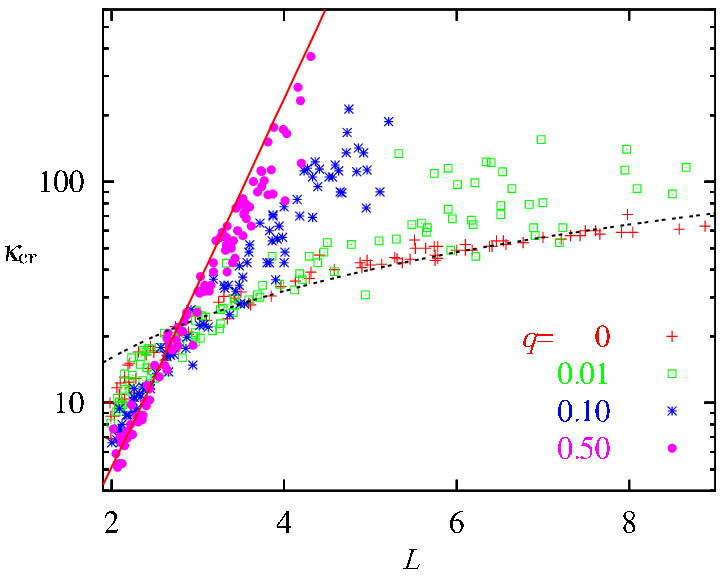}} \caption{(color online)
Dependence of the entrainment threshold on the depth $L$ for an ensemble
of small-world networks with $N=100,k=3$ and different randomness
$q$. The solid line is the same exponential function as that fitted to
the data with $pN=6(=2k)$ in Fig.~2. The dotted line is linear fitting
for the regular lattice ($q=0$).}  \label{fig:Kc-sw}
\end{figure}

The exponential dependence (\ref{exponential}) can be approximately
derived for asymmetric random networks with large $N$ and $pN$. In the
large-size limit, random graphs have locally a tree-like structure
\cite{dorogovtsev03}. The global tree approximation has previously been
used for determining statistical properties of random networks
\cite{newman01}. We apply here the same approximation and assume that
the graph of forward connections extending from the pacemaker node
represents a tree, so that any oscillator has only one incoming
connection from the previous shell. Then the shell populations $N_h$ are
given by $N_h=N_1 (pN)^{h-1}$ for $h=2,\ldots,H$, where $H$ is the total
number of shells determined by $\sum_{h=1}^{H} {N}_{h}=N$. Because $pN$
is large, we have $N_{h} \ll N_H \simeq N$ for $h<H$, and thus $L \simeq
H$.  Next, we estimate the numbers $m_{hk}$ of incoming connections
leading from all elements in the $k$th shell to an oscillator in the
$h$th shell. By definition of hierarchical shells, $m_{hk}=0$ if
$k<h-1$. In the tree approximation, $m_{hk}=1$ for $k=h-1$.  Because
most of the population is concentrated in the last shell, reverse
connections from other shells can be neglected. On the average, the
number of reverse connections from the shell $H$ to an oscillator in the
shell $h$ is $m_{hH}=p N_H$. Moreover, the relative statistical
deviation from this average is of order $(pN)^{-1/2}$, and is thus
negligible. Therefore, in this approximation all oscillators inside a
particular shell have effectively the same number of connections from
other shells, and a state with phase synchronization inside each shell
is possible. In this state, all oscillators inside a shell have the same
phase, i.e. $\phi_i=\theta _{h}$ for all oscillators $i$ in a shell
$h$. Under entrainment, the phases of such a state can be found
analytically as a solution of algebraic equations
\begin{eqnarray}
&-&\frac{\kappa }{pN}\sum_{k=h-1}^{H}m_{hk} -\sin (\theta _{h}-\theta
_{k})=1\quad \mbox{for $h=2,\ldots,n$},  \nonumber \\
&-&\mu \sin (\theta _{1}-\theta _{0})-\frac{\kappa }{pN}
\sum_{k=2}^{H}m_{1k}\sin (\theta _{1}-\theta _{k})=1,  \label{1}
\end{eqnarray}
where $\theta_0 \equiv \phi_0$. For large $pN$, we can linearize
$\sin(\theta_h-\theta_k)$ for $h,k \ge 2$ in the solution of
Eqs.~(\ref{1}) [it can be shown that $\theta_2-\theta_H$ is of order
$O(1/pN)$]. Furthermore, using that $N_H \simeq N \gg
N_{h}$ for $h<H$ and $L \simeq H$, we obtain for $h \ge 2$ that
\begin{equation}
 \sin(\theta_{h-1} - \theta_h) = \frac{pN}{\kappa} (1+pN)^{L-h}.
  \label{phase}
\end{equation}
Equations (4) determine the phases of oscillators in the considered synchronized state. Note that the explicit value of the phase $\theta_1$ in this state is not needed below. 

The entrainment breakdown can, in principle, occur through
destabilization of the synchronized state. Though the analytical proof
of its stability is not yet available, our numerical simulations show
that the synchronous entrained state with $0<\theta _{h-1}-\theta
_{h}<\pi /2$ is always stable when it exists. Thus, the breakdown of
entrainment in the considered system takes place in a saddle-node
bifurcation, through the disappearance of solutions of Eqs.~(\ref{1}).
This occurs when $|\sin(\theta_{h-1}-\theta_h)|=1$ for certain $h$.
For large enough $\mu$, we always have $0<\sin(\theta_0-\theta_1)<1$ (a
sufficient condition is $\mu >1+\kappa$). Among the other terms, the
term $\sin(\theta_1 - \theta_2)$ is always the largest one. Therefore,
the solution disappears and breakdown occurs when $\sin (\theta
_{1}-\theta _{2})=1$. Substituting Eq.~(\ref{phase}) into this equation
and solving it with respect to $\kappa$, we finally derive the
dependence (\ref{exponential}). Thus, we see that the entrainment
breakdown occurs through the loss of frequency locking between the first
shell and the rest of the network. The analytical estimate for the
critical coupling strength, obtained using the tree approximation,
agrees well with the numerical data, even for the networks which are not
very large.

So far we have used the coupling strength which was
rescaled as $\kappa \rightarrow \kappa /\Delta \omega $. Therefore, if
the non-scaled coupling strength is fixed, Eq.~(\ref{exponential})
determines the maximum $\Delta \omega_{\rm c}$ at which the entrainment
is still possible, $\Delta \omega_{\rm c} \propto \kappa
(1+pN)^{-L}$. The entrainment by a pacemaker can take place only if its
frequency lies inside the interval $(\omega ,\omega +\Delta \omega_{\rm
c})$. 

Thus, the entrainment window {\em decreases exponentially} with
the depth of a network. This is the principal result of our study, which
holds not only for standard random networks, where the above analytical
estimate is available, but also for small-world graphs and other
numerically investigated random topologies. In practice, it implies that
only shallow random networks with small depths are susceptible to
frequency entrainment. 

Our results remain valid when, instead of a pacemaker, external
periodic forcing acts on a subset of elements. We have checked that the
reported strong dependence on the network depth remains valid for
systems with larger network sizes, heterogeneity in frequencies of
individual oscillators, and several other coupling functions. The study
was performed for coupled phase oscillators which serve as an
approximation for various real oscillator systems, including neural
networks (see, e.g., \cite {kuramoto84,kuramoto91,kori}). Its
conclusions should be applicable for a broad class of oscillator
networks with random architectures.

Financial support by the Japan Society for Promotion of Science (JSPS) is
gratefully acknowledged.

\end{document}